# Possible Lattice and Charge Order in $Cu_xBi_2Te_2Se$

Yanan Li, Nathaniel P. Smith, William Rexhausen, Marvin A Schofield, and Prasenjit Guptasarma*

Department of Physics, University of Wisconsin - Milwaukee, Wisconsin, USA

E-mail: pg@uwm.edu

**Abstract**

Metal intercalation into layered topological insulator materials such as the binary chalcogenide $Bi_2X_3$ (X=Te or Se) has yielded novel two-dimensional (2D) electron-gas physics, phase transitions to superconductivity, as well as interesting magnetic ground states. Of recent interest is the intercalation-driven interplay between lattice distortions, density wave ordering, and the emergence of new phenomena in the vicinity of instabilities induced by intercalation. Here, we examine the effects of Cu-intercalation on the ternary chalcogenide $Bi_2Te_2Se$. We report the discovery, in $Cu_{0.3}Bi_2Te_2Se$, of a periodic lattice distortion (PLD) at room temperature, together with a charge density wave (CDW) transition around $T_d = 220K$. We also report, for the first time, a complete study of the $Cu_xBi_2Te_2Se$ system, and the effect of Cu-intercalation on crystal structure, phonon structure, and electronic properties for $0.0 \leq x \leq 0.5$. Our electron diffraction studies reveal strong Bragg spots at reciprocal lattice positions forbidden by ABC stacking, possibly resulting from stacking faults, or a superlattice. The c-axis lattice parameter varies monotonically with x for $0 < x < 0.2$, but drops precipitously for higher x. Similarly, Raman phonon modes $A_{1g}^2$ and $E_g^2$ soften monotonically for $0 < x < 0.2$ but harden sharply for $x > 0.2$. This indicates that Cu likely intercalates up to $x \sim 0.2$, followed by partial site-substitutions at higher values. The resulting strain makes the $0.2 \leq x \leq 0.3$ region susceptible to instabilities and distortions. Our results point toward the presence of an incommensurate CDW above $T_d = 220$ K. This work strengthens prevalent thought that intercalation contributes significantly to instabilities in the lattice and charge degrees of freedom in layered chalcogenides. Further work is required to uncover additional density wave transitions, and possible ground states such as superconductivity.

## 1. Introduction

Quantum Materials such as Bi-based topological insulators (TI) continue to generate interest among condensed matter physicists due to their non-trivial electronic structure, and possible applications as materials for quantum computing and spintronics [1-3]. Intercalation of Cu in layered $Bi_2Se_3$ (BS) and $Bi_2Te_3$ (BT) yields superconductivity [4,5] and, in the case of BS, a charge density wave (CDW) transition [6,7]. Previous studies of $Cu_xBi_2Se_3$ (CBS) and $Cu_xBi_2Te_3$ (CBT) have revealed that phase transitions in these systems are driven by the total content, nature of ordering, and location of Cu [8,9]. Electron diffraction studies of $M_xBi_2Se_3$ (M = intercalant metal) with higher concentration of the intercalant have revealed the formation of a superlattice (for M = Cu, Ag, Co, Sn, Fe) with a concomitant CDW, or an incommensurate CDW (I-CDW), ground state (M = Cu, Ag, Co) [10]. At these higher levels of intercalation, selected area electron diffraction (SAED) patterns display alternating bright and dark diffraction spots in the (001) zone axis [7]; these dark spots are forbidden in the ABC stacking of the host $Bi_2Se_3$. It is yet unclear whether the appearance of this superlattice pattern arises from symmetry breaking caused by stacking faults, or from a periodic feature caused by intercalation. Koski and colleagues have reported the observation of CDWs in intercalated $Bi_2Se_3$ based on electron diffraction results [6,7,10]. In their 1974 paper [11], Wilson, diSalvo and Mahajan had concluded from a study of large numbers of dichalcogenides that the adoption of a superlattice distortion usually foretells the possibility of a CDW. This, in turn, can co-occur with superconductivity. The question of whether an electronic modulation causes a structural distortion, or vice versa, is still a matter of debate in the literature [12]. In addition, the presence of such lattice and charge instabilities in the proximity of superconducting phases, such as the one in the binary chalcogenide, $Cu_xBi_2Se_3$ [4,13], remains of much interest in our efforts to clarify the microscopic origin of superconductivity in these systems. In this paper, we report the discovery of a superlattice-induced distortion and a charge density wave (CDW) transition in the Cu-intercalated ternary chalcogenide $Bi_2Te_2Se$ (BTS). We also report, for the first time, a complete study of $Cu_xBi_2Te_2Se$ ($0.0 \leq x \leq 0.5$), and the effect of Cu-intercalation on crystal structure, phonon structure, and electronic properties.

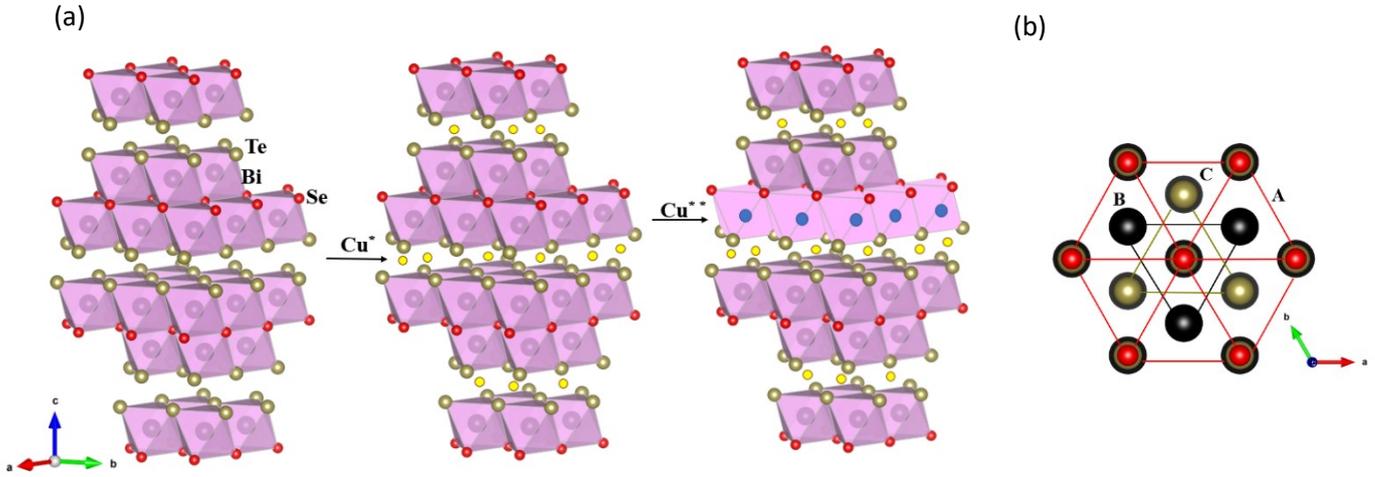

**Figure 1.** Crystal structure. (a) Crystal structure of pure $Bi_2Te_2Se$ (left), Cu intercalated $Bi_2Te_2Se$ (middle) and Cu substituted $Bi_2Te_2Se$ (right), and (b) with ABC stacking image as viewed looking down along the c-axis.

The topological insulator $Bi_2Te_2Se$ (BTS) is isostructural with $Bi_2Te_3$ and $Bi_2Se_3$, but with a better ordered structure, and larger bulk resistivity [14-16]. It has been well-known for decades [17] that $Bi_2(Te_{1-x}Se_x)_3$ is most ordered for $x = 1/3$ (which is $Bi_2Te_2Se$), due to the suppression of antisite defects, and Se/Te randomness, found in binary chalcogenides. This increased order is believed to be driven by stronger Se-Bi bonding. In spite of this, and in spite of the observation of superconducting order in intercalated BS and BT, intercalated BTS has not been explored. Here, we report the effects of introducing Cu ordering on the lattice and charge degrees of freedom by studying structural, vibrational, and electronic properties. Our results from x-ray diffraction, electron diffraction, Raman spectroscopy, and resistivity indicate that single crystal $Cu_xBi_2Te_2Se$ undergoes lattice expansion with increasing Cu, reaching full intercalation at Cu concentration of 20%. Higher Cu concentration reveals the presence of charge order and CDW transitions near $T_d \sim 220$ K. This is consistent with previous studies on layered binary chalcogenides (or dichalcogenides) such as $Cu$-$Bi_2Se_3$ [4], $4Hb$-$TaS_2$ and $1T$-$TaSe_2$ [11,12].

As shown in Figure 1(a) and 1(b), $Bi_2Te_2Se$ has a quintuple layer structure which forms as stacked Te-Bi-Se-Bi-Te. Five atomic layers are covalently bonded to form the so-called "quintuple layer," while adjacent quintuple layers (QLs) form intercalating spaces bonded via *van der Waals* interactions [16]. $Bi_2Te_2Se$ belongs to the $R\bar{3}m$ ($D_{3d}^5$) space group with a rhombohedral crystal structure. As in the case of $Bi_2Se_3$ and $Bi_2Te_3$, introduction of Cu during flux growth can result in Cu entering $Bi_2Te_2Se$ as either an intercalant in the *van der Waals* gaps, or as a substitutional defect at Bi sites. As an intercalant, $Cu^{1+}$ acts as a donor. As a Bi-site substituent forming a sigma bond, three Bi 6p electrons are replaced by one Cu 4s electron, resulting in a bi-valent amphoteric (ambipolar) defect.

## 2. Experimental Details

Single crystals of $Cu_xBi_2Te_2Se$ (x=0, 0.08, 0.12, 0.15, 0.2, 0.3, 0.5) were prepared by melting high-purity (99.999%) powders of Bi, Te, Se and Cu in stoichiometric ratios.

Stoichiometric mixtures of 2.5g batches were sealed into high-purity quartz tubes in vacuum after being weighed and sealed in an inert glove box, taking care to never expose to air. The crystals were grown from a melt using a two-step process. The mixtures in sealed quartz tubes were heated up to 850 C and maintained at that temperature for 48 hours. They were then cooled to 450 C at 0.1 C/min, followed by cooling to room temperature at 0.8 C/min. The tubes were subsequently re-sealed under vacuum and re-annealed at 600 C for two weeks before being quenched into tap water at room temperature, yielding high-quality single crystals.

Powder X-ray Diffraction (XRD) measurements were performed using a Bruker D8 Discover x-ray diffractometer with Cu Kα radiation. Powder X-ray diffraction data were collected from pieces of single crystals powdered inside an inert glove box. Rietveld refinement was performed using GSAS (General Structure Analysis System) and the EXPGUI interface. Raman spectroscopy measurements were performed on a Renishaw Inc. 1000B, 1800 grating Raman spectrometer equipped with a microscope. Samples were measured at room temperature using an excitation wavelength of 632 nm through a 20x microscope objective lens, resulting in a laser spot of 4 $\mu m^2$ with power of 35 mW at 100%. Data presented here were collected using 1% power (0.0875 mW/$\mu m^2$ and 10% power (0.875 mW/$\mu m^2$), while keeping the same microscope objective and laser spot size. The polarization configuration used in our measurements was: Z(XX)$\bar{Z}$. Phonon peaks were fitted as Lorentzian functions.

X-ray Photoelectron Spectroscopy (XPS) was performed with an Mg anode source in a Perkin Elmer PHI 5440 ESCA System in ultra-high vacuum. Surfaces were prepared by cleaving the crystal surface in inert atmosphere. However, we believe that the transfer process resulted in some exposure to air. For this reason, we studied the Cu oxidation state using depth profiling by ablating with an Ar-ion gun. XPS was then used to study the oxidation state of Cu with increasing ablation of the top layer. XPS spectra up to 1000 eV were fitted using a Lorentzian fitting function convolved with a Gaussian after subtracting the baseline. Selected Area Diffraction (SAED) was performed at room temperature with a Hitachi H-9000NAR high resolution transmission electron microscope (HRTEM) operating at 300kV. For this, crystals of $Cu_xBi_2Te_2Se$ (0 ≤ x ≤ 0.5) were mechanically ground in an inert glove box and dispersed on to Lacey-carbon grids. Variable temperature resistivity studies down to 1.8 K were performed using 4-probe silver paste contacts on single crystals surfaces and placed in a Quantum Design Physical Property Measurement System (PPMS) equipped with a 9T magnet.

## 3. Results and Discussion

### 3.1 X-ray Diffraction

Figure 2(a) shows results of X-ray Diffraction and analysis on powdered single crystal $Cu_xBi_2Te_2Se$ for x=0.00, 0.08, 0.12, 0.20, 0.30 and 0.50. While the results in Figure 2(b) are from Rietveld refinement of the entire diffractograms, the inset in Figure 2(a) is shown as a visual indication that the (006) reflection follows the c-axis changes as described and reported in Figure 2(b). As seen in Figure 2(a), the c-axis value increases substantially with increasing Cu content, indicating intercalation into the van der Waals gap. Thus, Cu intercalation results in the c-axis increasing from c = 29.94 Å (for x = 0, the parent phase) to c = 30.08 Å (for x = 0.20). With higher concentration, x=0.30 and x=0.50, the refinement shows c-axis dropping to 29.97 Å.

Previous work [4,5] shows that Cu in $Cu_xBi_2Se_3$ or $Cu_xBi_2Te_3$ may either intercalate between the Se/Te layers or substitute for Bi within the host structure. Using a combination of results from XRD and Raman spectroscopy (described later in this section), we suggest that most of the Cu in $Cu_xBi_2Te_2Se$ gets intercalated into the gap for x < 0.3, but that higher Cu levels (x ≥ 0.3) results in partial Cu substitution at Bi sites. Thus, for x ≤ 0.2, the presence of Cu in the van der Waals spaces between the quintuple layers (QL) results in an increase in c-axis length. However, if additional Cu partially substitutes at Bi sites for x ≥ 0.3, then a smaller $Cu^{2+}$ ion (ionic radius = 0.72 Å) replacing a larger $Bi^{3+}$ ion (ionic radius = 1.08 Å) could lead to a decrease in the c-axis parameter. Although we do not completely understand the reasons for the changes observed in lattice parameters and phonons, we wish to highlight the fact that there are significant crystal structure- and phonon-related changes in the 0.2 ≤ x ≤ 0.3 region.

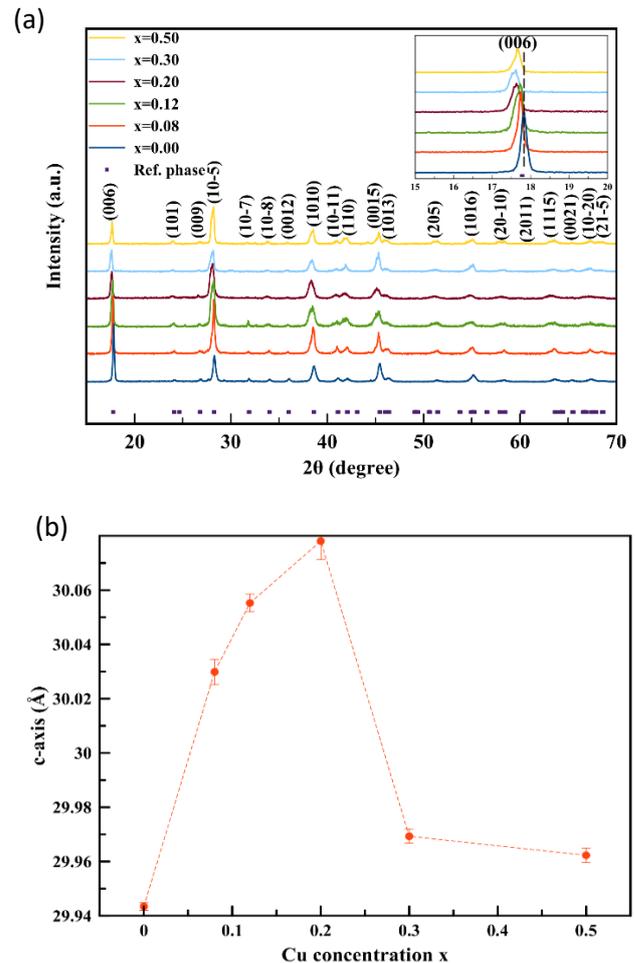

**Figure 2.** (a) Power X-ray Diffraction (XRD) patterns of the as-grown $Cu_xBi_2Te_2Se$ single crystals. Inset: the (006) peak shifting with increasing Cu concentration x [bottom to top, x=0.00, 0.08, 0.12, 0.2, 0.3 and 0.5]; (b) Rietveld refined c-axis values shifting with Cu concentration x.

### 3.2 Selected Area Electron Diffraction

Transmission Electron Microscopy (TEM) and Selected Area Electron Diffraction (SAED) studies were carried out in order to further assess the crystallinity and morphology of $Cu_xBi_2Te_2Se$ samples. Figure 3 is a typical bright-field TEM image (Figure 3(a)) of x=0.3 $Cu_xBi_2Te_2Se$ flake and its corresponding SAED pattern (Figure 3(b)) obtained from the same area. As indicated in Figure 3(b), the flake is oriented close to the <001> zone-axis and shows good crystallinity, consistent with strong diffraction contrast apparent in the bright-field image of Figure 3(a). Flakes prepared for TEM studies (mechanically ground and dispersed on Lacey-carbon grid) showed large micron-sized regions, and a tendency for a layered morphology, as evident in Figure 3(a). Studies of $Cu_xBi_2Te_2Se$ samples for x=0, 0.12, 0.2, 0.3 and 0.50 were performed. The general results shown in Figure 3(a) are typical of all samples examined.

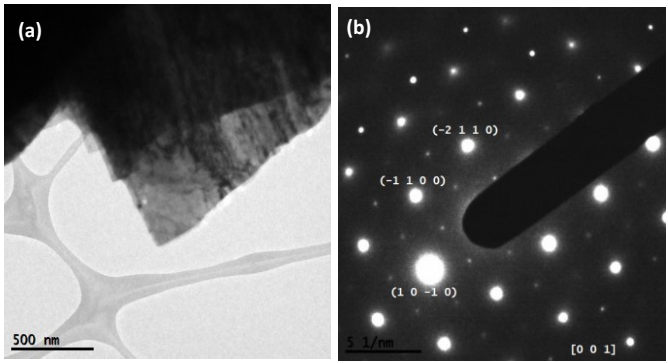

**Figure 3.** Selected Area Electron Diffraction (SAED). (a) bright-field TEM image of an x=0.3 $Cu_xBi_2Te_2Se$ flake, (b) corresponding SAED pattern obtained from the same area in (a).

In Figure 3(b), the (100) and (010) reflections, which are kinematically forbidden for ABC stacking in the rhombohedral parent phase of $Bi_2Te_2Se$ [18], are clearly evident as weak reflections. The appearance of such spots could indicate that either the c-axis ABC stacking sequence symmetry is broken, or that a superlattice ($\sqrt{3}a \times \sqrt{3}a$, $R = 30°$) reconstruction in the ab-planes is retaining the c-axis stacking order [7]. The intercalated Cu lie mostly in interstitial sites in the van der Waals gap, and it is reasonable to suppose that this effect might cause a disruption of the ABC stacking order and appearance of the kinematically forbidden reflections, as observed and concluded in prior studies of Sn, Fe, Co, or Cu- intercalated single crystals of the $Bi_2Se_3$ single crystals [10].

It is important to examine whether the forbidden diffraction spots that appear in Fig. 3(b) for x=0.3 $Cu_xBi_2Te_2Se$ are due to dynamic scattering, especially considering the strong diffracting conditions of the <001> zone-axis orientation. To address this question, the sample was tilted off the <001> zone-axis in order to reduce dynamical scattering. Figure 4(a) is an SAED pattern obtained from the same region shown in Figure 3(b). In spite of the tilt, the SAED reveals a diffraction pattern clearly associated with the same forbidden reflections. For direct comparison, Figure 4(b) is an SAED pattern from our x=0 $Cu_xBi_2Te_2Se$ sample that is similarly tilted off the <001> zone-axis. However, the forbidden reflections remain. We conclude from this that the pure x=0 sample likely supports some degree of stacking faults based on the layered nature of the material, either intrinsic to the crystal growth method or introduced during TEM sample preparation through the mechanical grinding steps necessary to reduce the bulk crystal to sufficiently small sizes for TEM studies. In both cases (x=0 and x=0.3), the presence of forbidden (100) and (010) reflections remain comparatively strong when the samples are tilted off the <001> zone-axis with an intent to reduce dynamic scattering effects. This would be the case if the reflections are due to a 2D scattering mechanism where the diffraction spots are rods in reciprocal space and is consistent with the picture of faults in the ABC stacking symmetry.

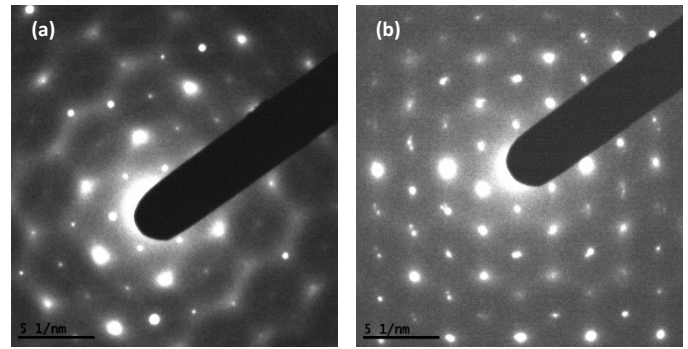

**Figure 4.** SAED patterns recorded tilted slightly off the <001> zone axis for (a) x=0.3 and (b) x=0.0.

Note also, in Figure 4(a), evidence of diffuse scattering structure in the SAED pattern from the x=0.3 crystal. Similar features are weak or non-existent in Figure 4(b), from x=0. The diffuse scattering features shown for the x=0.3 $Cu_xBi_2Te_2Se$ sample (Figure 4(a)) appear directly related to Cu content; those features are essentially absent in the pure x=0 sample. Similar diffuse features in SAED patterns have been observed on other layered dichalcogenides where the authors have concluded that the formation of a superlattice at a certain higher temperature foreshadows the possibility of a charge density wave transition ($T_d$) at lower temperature [11]. The authors further concluded that the satellite spots above $T_d$ arise from an incommensurate CDW. We argue below that our SAED observations (performed at room-temperature) are indications of the formation of an induced symmetry-breaking for x=0.3 $Cu_xBi_2Te_2Se$, at the edge of the regimes between intercalation and substitution, and that these results are consistent with the presence of a CDW transition at lower temperature observed in resistivity measurements, and shown in Figure 6.

Y Li *et al*

To further examine the effect of Cu on crystal structure distortion, we show in Figure 5 our results of SAED studies of $Cu_xBi_2Te_2Se$ samples for x = 0, 0.12, 0.2, 0.3 and 0.50. Figures 5(a) through 5(f) are SAED images recorded with the beam aligned closely along the <001> zone axis ("on-axis"). Figures 5(a′) through 5(f′) show SAED recorded on the same flake, at the same spot, but with a slight tilt from the <001> zone axis ("off-axis"). On-axis diffraction patterns (Figure 5 (a)-(f)) indicate good crystallinity, and confirm that the crystals are, on average, uniform and well-ordered. Tilting slightly off the zone axis helps minimize the intensity of the Bragg reflections to enable the observation of weak intensities between Bragg reflections. A comparison of the left and right panels in Figure 5 shows that weak, diffuse intensities show up, off-axis, for all samples. These are especially evident for x = 0.2 and x = 0.3, and are highlighted with small arrows in yellow. As discussed below, we interpret the observation of weak intensities ("streaks") between Bragg reflections to mean that an incommensurate charge order develops in BTS. We find that the streaks intensify with increasing Cu concentration in the region $0.2 \leq x \leq 0.3$, correlated with distortion and phonon changes observed in the same region, Fig 2(b) and Fig 9(b) and (c), serving to help intensify the I-CDW.

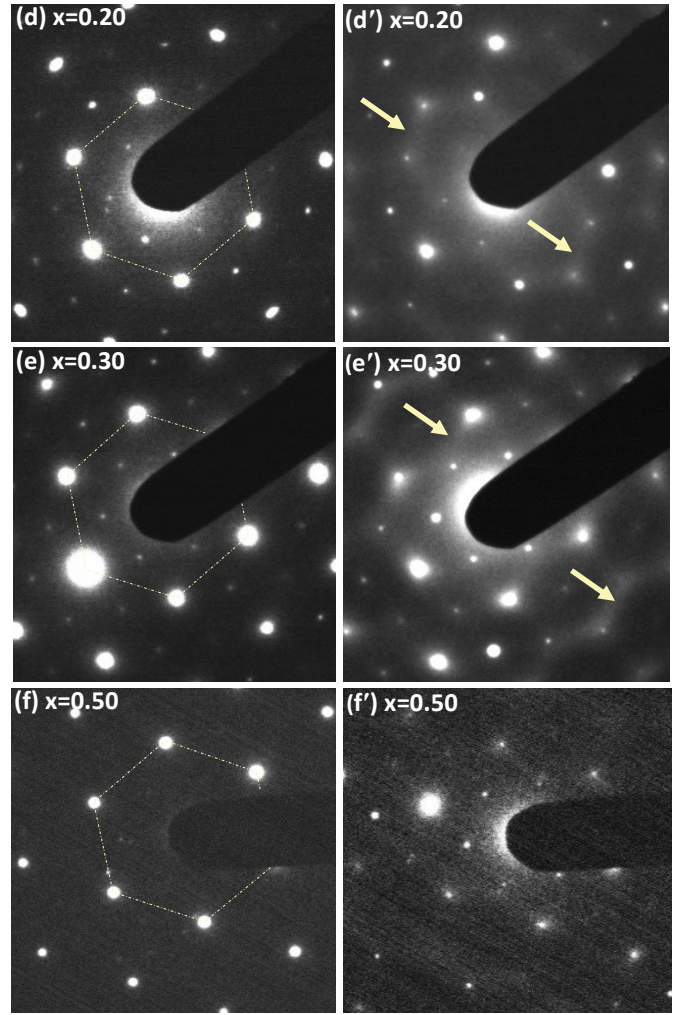

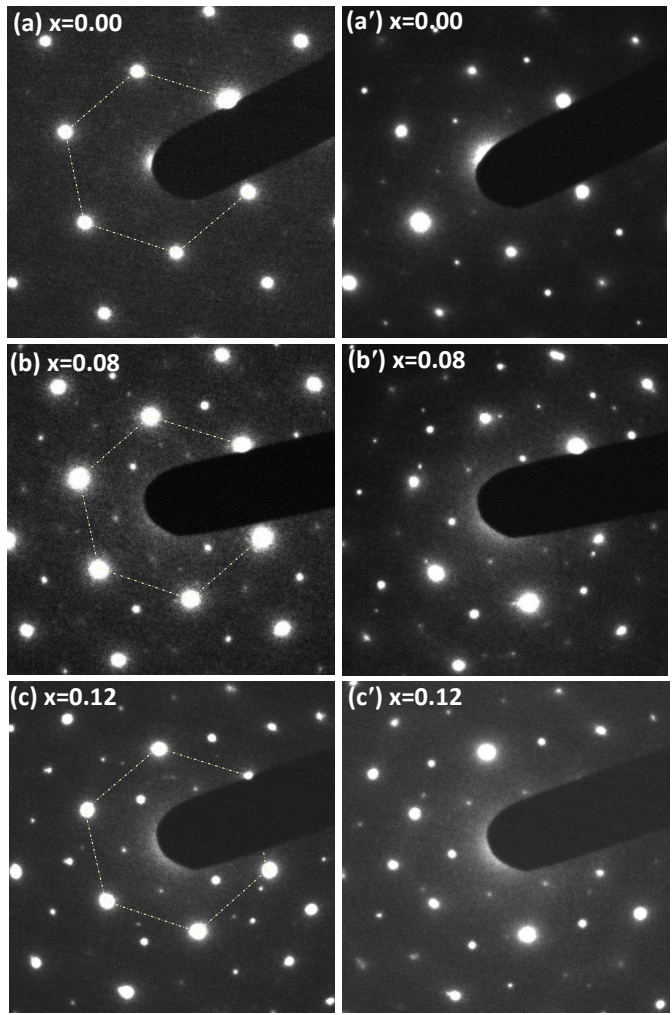

**Figure 5.** Comparison of SAED for as-grown samples (x=0.00 to 0.50). (a)-(f) SEAD recorded close to <001> zone axis, or "on-axis"; (a′)-(f′) same diffraction area and spots as (a)-(f) but slightly tilted off the <001> zone axis, or "off-axis." Yellow arrows indicate streaks (diffuse regions) which intensify for certain values of x. We have optimized contrast to help the reader visualize reflections more clearly.

A CDW is often found to co-occur with a Periodic Lattice Distortion (PLD) [12] such that the periodicity of the CDW is commensurate with the periodicity of the underlying atomic lattice distorted by the PLD. The electronic charge density, a scalar quantity and a natural order parameter of a CDW, is normally assumed to be linearly coupled to the longitudinal PLD, with the ordered CDW period an integral or fractional multiple of the period of the atomic lattice or the PLD. Such a situation could also arise due to planar defects such as stacking faults [20][21], as is likely to be the case in our samples, given the large changes in lattice constants and phonon frequencies in the $0.2 \leq x \leq 0.3$ region.

Let us now consider the effect of disorder on CDW periodicity such that the periodicity is not represented in the reciprocal lattice by specific vectors $\vec{k}$, but rather one that exists over an entire range of vectors $\vec{k}\pm\Delta\vec{k}$, on either side of $\vec{k}$. This would then lead to diffraction patterns with "diffuse" intensity centered around $\vec{k}$. In the case of large $\Delta\vec{k}$, it is possible to have a scenario in which diffraction patterns appear as diffuse streaks between Bragg reflections, with the CDW maintaining the overall symmetry of the underlying lattice but not commensurate with the lattice periodicity. This is the basis of the idea of an incommensurate CDW, or I-CDW [19].

In an effort to minimize strain energy, cooling below a certain temperature $T_d$ could cause an incommensurate CDW (I-CDW) to lock-in to a period commensurate with the period of the lattice or the PLD. Such a transition from an I-CDW to a CDW can be likened to a temperature-driven disorder-to-order transition. An I-CDW, usually observed in diffraction patters as diffuse patterns and streaks, acts as a precursor to a CDW [11, 22-25]. SAED studies of 1T-TaSe$_2$, 2H-NbSe$_2$, 1T-(Ta$_{0.6}$Nb$_{0.4}$)S$_2$ and 1T-Ta$_x$Ti$_{1-x}$S$_2$ have shown marked diffuse scattering for T > $T_d$, but sharp diffraction spots below $T_d$ [22][23][25][19]. In the case of Bi$_2$Se$_3$, addition of Cu/Ag/Co/Fe at high concentration yields satellite diffraction spots, interpreted by Koski *et al* to be a signature of an I-CDW [10]. Based on the above, we conclude that in our samples, the observation of an I-CDW at room temperature in Figure 5(e′) foreshadows the transition to the CDW observed in Figures 6(a) and 6(b) near 200 K.

Note that some weak diffuse intensity is also seen in other samples, as is clear upon careful examination of Figures 5 (b′, c′, d′, e′, j′ and f′) and Fig 4(b). The tendency for diffraction intensity arising from an underlying I-CDW seems to be an intrinsic property of the BTS system [21]. Our results point toward Bi$_2$Te$_2$Se being intrinsically disordered due to Se/Te dislocation, or due to intercalation by stray Bi atoms inhabiting intercalating spaces [26]. Thus, a weak underlying I-CDW in BTS, already in existence, is enhanced by additional distortions introduced by Cu, as demonstrated in Figures 2 and 9.

### 3.3 Resistivity with varying Temperature

Figure 6(a) shows 4-probe resistivity measurements on the x=0.3 single crystal as a function of temperature between 2K and 300K. Note two hump-like features near 220K and 255 K, signaling metal-to-insulator like transitions reminiscent of transitions resulting from a Charge Density wave. The inset to Figure 6(a) shows a plot of dρ/dT as a function of temperature, developing two separate minima near 220K and 255K at the onset of each transition. Figure 6(b) displays ρ(T)/ ρ(300K) as a function of temperature for as-grown single crystal with x=0.00, 0.12, 0.20 and 0.30. Metallic behavior was seen on samples with x=0.00, 0.12 and 0.2. At x = 0.5, excessive Cu doping appears to deteriorate the quality of the crystal as grown. For x = 0.5, we were unable to obtain large enough single crystals for 4-probe resistivity measurements. On the other hand, powder diffraction SAED, and Raman spectroscopy do not require large crystals; these results are presented in Figures 2, 5, and 9.

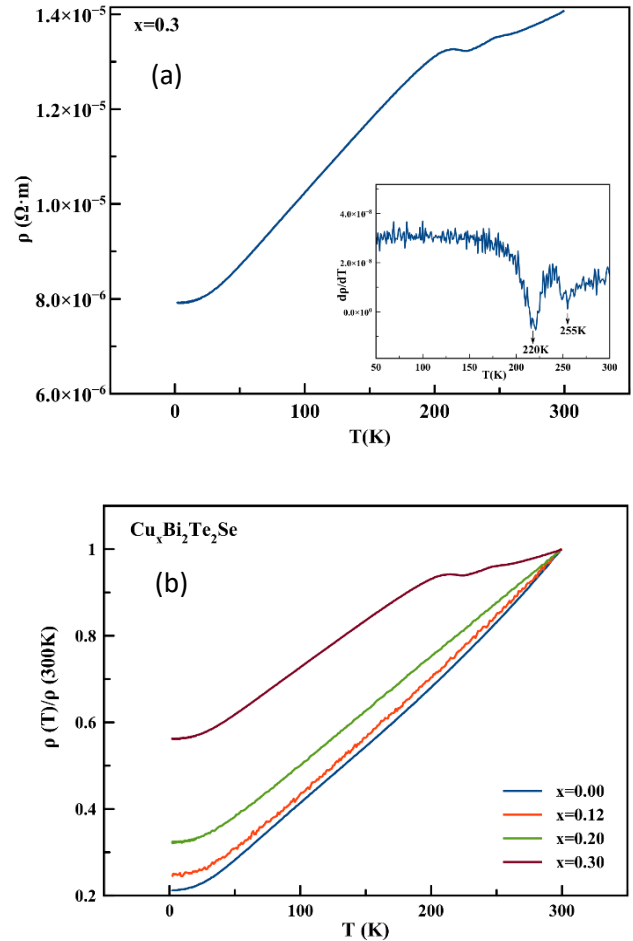

**Figure 6.** (a) Resistivity vs temperature (2 - 300K) on Cu$_{0.3}$Bi$_2$Te$_2$Se. Inset: dρ/dT vs T curve enlarged from 50K to 300K. Arrows indicate the minimum point of dρ/dT, indicating a Charge Density Wave (CDW) transition onset temperatures. (b) ρ(T)/ ρ(300K) vs temperature (2-300K) for x=0.00, 0.12, 0.20 and 0.30 samples.

### 3.4 X-ray Photoelectron Spectra

One of the questions about integrating Cu into the sample during crystal growth is whether Cu actually intercalates into the van der Waals layers or substitutes into the main lattice. Intercalated Cu has been shown to have oxidation states of either 0 or 1+, whereas substitution into the Bi site would yield Cu$^{2+}$. Figure 7 shows XPS results from the surfaces of x=0.00 and x=0.2 samples, revealing signals from Bi 4f, Te 3d, Se 3d, C 1s and O 1s. Untreated Cu$_x$Bi$_2$Te$_2$Se samples show features for both Cu$^{1+}$ and Cu$^{2+}$ oxidation states in Cu 2p. At this stage, it is important to rule out the presence of opportunistic atoms of oxygen which could yield Cu$^{2+}$ due to the formation of CuO. As discussed below, we find clear evidence for the presence of only Cu$^{1+}$ after ablating approximately 50 nm from the crystal surface. In Figure 7(a), before sputtering, the $Cu_{0.2}Bi_2Te_2Se$ sample reveals Cu 2p binding Energies (BE) at 932.27 and 952.35eV (corresponding to $Cu^{1+}$), and at 934.48 and 954.69eV (corresponding to $Cu^{2+}$) together with two strong satellite peaks [6] which are the characteristics of $Cu^{2+}$. After 5 minutes of Ar ion sputtering at 0.5kV, the Cu 2p peaks are found to be centered at 932.50 and 952.58eV. Thus, upon ablation, we observe the presence of nearly entirely $Cu^{1+}$, with the $Cu^{2+}$ feature almost gone, after an additional 3 minutes of 3 kV etching. The binding energy of Cu 2p is at 932.65 and 952.64eV, clearly demonstrating features of the $Cu^{1+}$ oxidation state. This result is consistent with the published literature for the intercalated dichalcogenide $Cu_xBi_2Se_3$ [27,28].

A comparison of BE for Bi 5d between x=0 and x=0.2 in $Cu_xBi_2Te_2Se$ samples is also shown in Figure 7(b), for samples after 0.5kV sputtering. The binding energies of Bi 4f 7/2 and Bi 4f 5/2 in $Bi_2Te_2Se$ (shown in red) are 157.47 eV and 162.78 eV, respectively. A slight shift to a lower BE can be observed for the $Cu_{0.2}Bi_2Te_2Se$ sample (shown in blue). This is further indication of Cu intercalation, in agreement with other reports for $Cu_xBi_2Se_3$ [9]. We conclude from this that, for $0 < x \leq 0.2$, Cu exists in the form of singly ionized interstitial atoms of $Cu^{1+}$ acting as donors and located in the intercalating van der Waals spaces [9,27]. Note that $Cu^{2+}$ would replace $Bi^{3+}$ sites by creating two holes which act as acceptors [27,28].

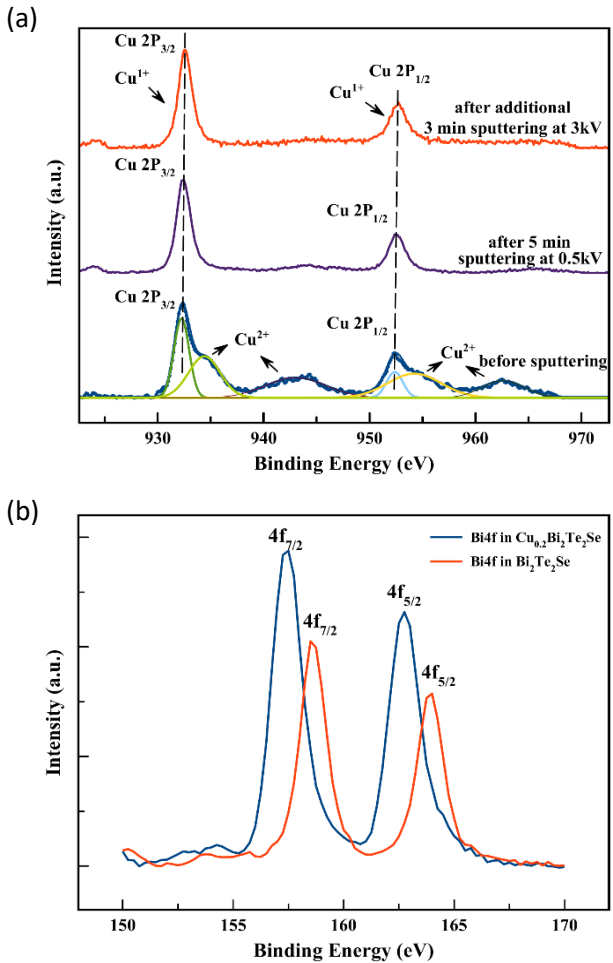

**Figure 7.** X-ray Photoelectron Spectra for the orbitals: Cu 2p and Bi 4f. (a) Comparison of Cu 2p for freshly cleaved $Cu_{0.2}Bi_2Te_2Se$ (bottom spectrum), followed by Ar ion gun 0.5kV 5-min surface sputtering (middle spectrum), followed again by additional 3kV 3-min sputtering (top spectrum). Before sputtering, both $Cu^{1+}$ and $Cu^{2+}$ are observed on the sample surface. The $Cu^{2+}$ satellite peaks are labeled with arrows. Note that, after sputtering, only $Cu^{1+}$ peaks are left. (b) Comparison of Bi 4f orbitals' shift between $Bi_2Te_2Se$ and $Cu_{0.2}Bi_2Te_2Se$. With Cu intercalation, both $4f_{5/2}$ and $4f_{7/2}$ peaks shift to the left of $Bi_2Te_2Se$.

### 3.5 Raman spectroscopy

We now examine the effect of Cu on phonon modes in $Cu_xBi_2Te_2Se$. Figure 8 displays our results observed on pure (x = 0) $Bi_2Te_2Se$. Consistent with the literature for Raman modes from $Bi_2Te_2Se$ [17, 29, 30] under $z(xx)\underline{z}$ polarization, we find the $A_{1g}^2$ mode (149cm$^{-1}$), the $E_g^2$ mode (104cm$^{-1}$), a splitting mode M (135cm$^{-1}$) of $A_{1g}^2$, and an extra mode "P" (115cm$^{-1}$), identified in the literature as possibly arising from $E_g^2$ mode splitting [17, 29]. Typically, there are four Raman-active modes for $Bi_2Te_2Se$, $A_{1g}^1$, $A_{1g}^2$, $E_g^1$ and $E_g^2$, where $A_{1g}^1$ and $E_g^1$ show up below 100cm$^{-1}$ [29]. Our filter begins to cut off intensity starting around 100cm$^{-1}$ – thus, our lower frequency modes $A_{1g}^1$ and $E_g^1$ (50cm$^{-1}$ to 80cm$^{-1}$) are indistinguishable from the noise. Our analysis is therefore focused on the behavior of the $E_g^2$ and $A_{1g}^2$ modes with varying x. The M mode, previously identified in the literature, is thought to be a local mode arising from Se/Te antisite defects [17, 29, 31]. The so-called P mode (115cm$^{-1}$) is thought to arise from antisite defects between Te and Se [17]. In this scenario, Se and Te in $Bi_2Te_{3-x}Se_x$ ($0 \leq x \leq 1$) can randomly replace each other, forming both Bi-Te and Bi-Se bonds [17]. These adjacent Bi-Te and Bi-Se bonds can decouple and lead to two-mode behavior in $A_{1g}^2$ (resulting in the M mode) and in $E_g^2$ (resulting in the P mode).

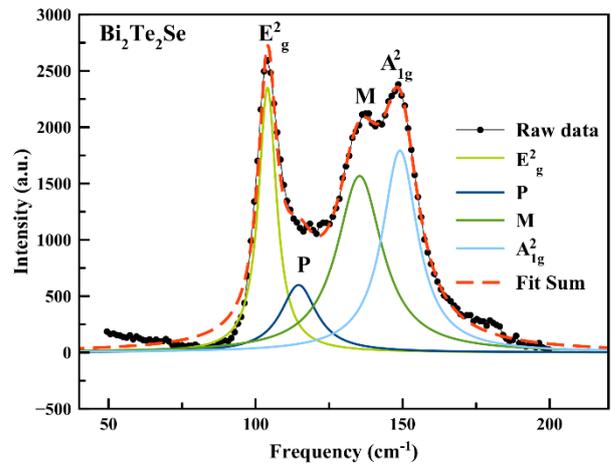

**Figure 8.** Raman spectra of $Bi_2Te_2Se$ taken at room temperature with 10% laser power under a $z(xx)z$ polarized configuration. Raw

data after baseline subtraction is shown by black dots. Lorentzian functions fitted with four individual modes are labeled as $E_g^2$, P, M, and $A_{1g}^2$.

Figure 9 (a)-(c) shows the effect of Cu in Bi$_2$Te$_2$Se for $A_{1g}^2$ and $E^2$g modes. Figure 9(a) is the raw data, shown here together with Lorentzian fits. Figure 9(b) and 9(c) show the variation of the peak positions of $A_{1g}^2$ and $E_g^2$ with total Cu concentration, x. To examine possible effects of local heating from the focused laser, Raman spectra were measured at both 1% power (0.0875 mW/μm$^2$) and 10% power (0.875 mW/μm$^2$), separately. As shown in Figure 9(b) and 9(c), peak shifts for the two different laser power settings were within each other's error bars. In Figure 9(b) and 9(c), $A_{1g}^2$ and $E_g^2$ phonons clearly soften with increasing Cu concentration between x=0.0 and x=0.2. However, with higher Cu concentration, of x=0.3 and x=0.5, we observe a hardening of the $A_{1g}^2$ and $E_g^2$ phonon modes, together with a broadening of the peak widths. Previous reports have ascribed similar softening versus hardening of A$_{1g}$ and E$_g$ modes to intercalation and substitution [32, 33]. Chen et al [33] conclude that the $A_{1g}^1$ mode in Bi$_2$Se$_3$ shifts to lower frequency with increasing Cu intercalation. In contrast, substitution of lighter atoms at the Bi site has been found to result in $A_{1g}^1$, $A_{1g}^2$, $E_g^1$ and $E_g^2$ modes shifting to higher frequency [17, 32].

We could try to understand this as follows. When Cu enters as an intercalant between two quintuple layers (QL), each QL experiences an extra Coulomb force resulting from the Cu in the van der Waals spaces [4, 9, 33]. This Coulomb interaction can modify bond-lengths and structure of the QL, thus affecting the phonon vibrational frequency. The hardening of phonon modes for x ≥ 0.3, compared with lower x ≤ 0.2, is in reasonable agreement with our XRD results. In other words, lower Cu concentrations lead to Cu intercalation whereas higher Cu concentrations can lead to partial substitution of Cu$^{2+}$ to Bi$^{3+}$. The increase in peak width for x=0.3 and x=0.5 is likely due to additional disorder introduced by the Cu intercalant when it substitutes into the Bi$_2$Te$_2$Se host structure [34]. We conclude from this that Cu incorporation in Cu$_x$Bi$_2$Te$_2$Se primarily occurs as an intercalant at low concentrations, as indicated by phonon modes shifting to lower frequencies when compared to pure Bi$_2$Te$_2$Se. At higher concentrations, Cu incorporates as a substituent, as indicated by phonon modes shifting to higher frequencies.

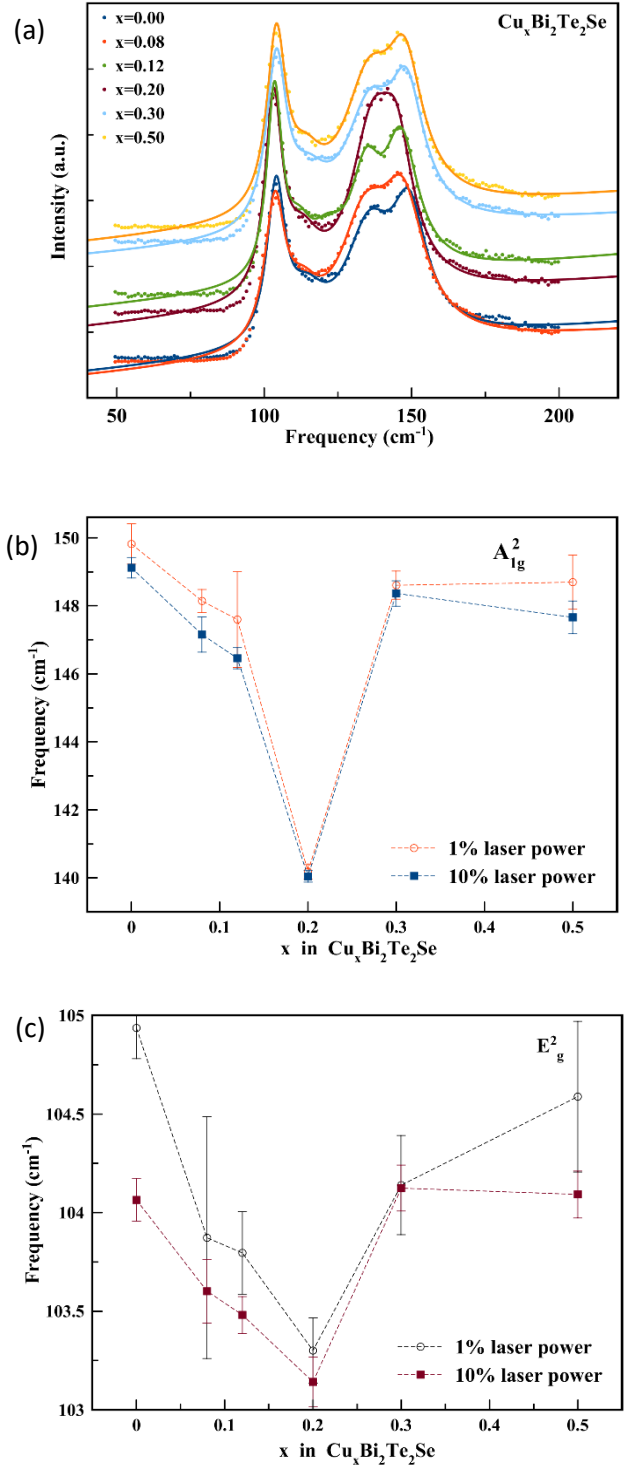

**Figure 9.** Raman spectra under z(xx)z̄ polarization. (a) Raman spectra of Cu$_x$Bi$_2$Te$_2$Se (from bottom to top, x=0.00, 0.08, 0.12, 0.2, 0.3 and 0.5) raw data with Lorentzian fit. (b) and (c) are $A_{1g}^2$ and $E_g^2$ peak dependence with Cu concentration x; Empty circles and solid squares in (b) and (c) are a comparison to ensure reproducibility between 0.0875 mW/μm$^2$ and 0.875 mW/μm$^2$ incident laser power.

**3.6 Further Discussions**

We now return to discuss the origin of the CDW observed in $Cu_{0.3}Bi_2Te_2Se$. A typical CDW phase transition is accompanied by the opening of a gap at the Fermi level, resulting in a metal-to-insulator like transition in resistivity as a function of temperature. In most microscopic models, a CDW transition is understood as being driven by either an electron energy instability near the Fermi level, or due to Fermi nesting [35-37]. Most commonly, CDW phase transitions are observed in 1D chains [38], or in 2D layered di-chalcogenide materials [39]. In practice, single crystals with *imperfect chains* (quasi-1D), or *imperfect nestings* (quasi-2D), can lead to such an energy gap being partially opened. This is more often seen as a metal-to-semimetal transition similar to our observations in Figure 6 in our data, and as also observed in multiple other systems [40, 41]. The CDW transition observed here in $Cu_{0.3}Bi_2Te_2Se$ is likely to be related to the nature of the quintuple layers in this material. As seen from X-ray diffraction, increasing Cu concentration lengthens the c-axis. The crystal lattice is likely to respond to this by forming new periodic structures, which we observe as a superlattice in electron diffraction results from our crystals. The lattice might also respond by creating stacking faults, or other disorder types, helping reveal forbidden reflections in the ABC stacking which we observe. At this point, it is difficult to tease out whether the forbidden reflections arise from a superlattice, or from other types of disorder. Based on the fact that the reflections are retained when the crystals are tilted from the (001) zone axis, we conclude that the forbidden reflections do not arise from dynamic scattering. Thus, a *periodic lattice distortion* (PLD), such as the ones described above, is often accompanied by a charge density wave (CDW) [12, 37, 42]. The "fundamental origin" question, of whether an electronic modulation causes a structural modulation, or vice versa, is still being debated in the current literature [12, 19, 42, 43]. We have demonstrated here the presence of both a possible PLD and a CDW in $Cu_xBi_2Te_2Se$, without commenting upon which one leads to the other.

The presence of diffuse scattering in Figure 5(e′) is clearly indicative of charge ordering. Most authors interpret such intensities in electron diffraction as arising from the formation of an incommensurate charge density wave (I-CDW). The combined observation of an incommensurate CDW in SAED at room temperature, together with the transitions observed in resistivity near 220-225 K, indicate that the feature in resistivity is a transition from an I-CDW to a (possible) CDW phase below 220 K. Rossnagel provides a review of both theoretical and experimental results for a series of dichalcogenides [12], and uses existing data to indicate that the physical picture provided by the Peierls model is likely to be correct for systems such as $1T-TaS_2$, $2H-TaS_2$ and $1T-TiSe_2$. An incommensurate CDW/PLD is mostly associated with long coherence length, small energy gap and weak electron-phonon coupling. Such a transition is driven by an instability of the Fermi surface, such as in $2H-TaS_2$ [11]. On the other hand, in the strong-coupling regime, a CDW shows up with short coherence, and a large energy gap. The latter case is driven by an ionic-covalent bonding picture similar to the case in $1T-TiS_2$ [44]. In the broader picture, however, it has not been easy to determine a strong dominant driving force for the origin of CDWs in layered materials: strong electron-coupling and an appropriate density of states near the Fermi level certainly seem to be important. Further work is needed in order to tease-out the details of the origin of CDW in these materials, and the possible discovery of ground states such as superconductivity.

**4. Conclusions**

In summary, we used a self-flux method to grow single crystals of $Cu_xBi_2Te_2Se$ and examined the effect of Cu on crystal structure, phonon and electron properties. X-ray photoelectron studies show that Cu exists in the intercalating spaces as $Cu^{1+}$. Increasing Cu concentration leads to an increase in the c-axis length up to x=0.2. Beyond x=0.2, the c-axis drops to lower values. Similarly, Raman modes $A_{1g}^2$ and $E^2{}_g$ soften up to x=0.2 and harden for higher values of Cu. This points to the possibility that, while lower concentrations of Cu for $x \leq 0.2$ end up with Cu in the intercalating spaces, higher concentrations end up as substituents. This conclusion is consistent with previous observations in $Cu_xBi_2Se_3$, and also makes the $0.2 \leq x \leq 0.3$ region in the $Cu_xBi_2Te_2Se$ phase diagram a region of much interest due to the presence of possible instabilities and distortions. The observation of a charge density wave (CDW) transition in the x=0.3 sample is therefore interesting. In particular, $Cu_{0.3}Bi_2Te_2Se$ reveals diffuse order in electron diffraction together with CDW-like transitions in resistivity near 220 K. In addition, we also find the existence of reflections forbidden by ABC stacking. These are indicative of a periodic lattice distortion (PLD), often observed as harbingers of a CDW transition in other layered di-chalcogenides. Our work provides strong indications that the Cu-intercalated ternary chalcogenide $Cu_{0.3}Bi_2Te_2Se$, upon cooling below 220K, undergoes a transition from an incommensurate charge density wave (I-CDW) to a charge density wave (CDW) state. Further work is needed, with temperature-dependent electron diffraction, in order to confirm this. Additionally, it is important to study this system further in order to uncover the possibility that an incommensurate CDW with a weak electron-phonon coupling could eventually lead to the discovery of superconductivity in $Cu_xBi_2Te_2Se$.

**Acknowledgements**

We thank Dr. Steven Hardcastle and Dr. Afsaneh Dorri of the University of Wisconsin Advanced Analysis Facility for their support during our measurements. We acknowledge prior funding from AFOSR-MURI which enabled this work. We also thank the Lichtman family for a research award and fellowship provided to YL.

**References**

[1] Fu L and Kane C L 2007 Topological insulators with

inversion symmetry *Phys. Rev. B - Condens. Matter Mater. Phys.* **76**, 1–17

[2] Hasan M Z and Kane C L 2010 Colloquium: Topological insulators *Rev. Mod. Phys.* **82**, 3045–3067.

[3] J. E. Moore 2010 The birth of topological insulators. *Nature* **464**, 194–8

[4] Hor Y S *et al.* 2010 Superconductivity in $Cu_xBi_2Se_3$ and its Implications for Pairing in the Undoped Topological Insulator. *Phys. Rev. B* **104**, 057001

[5] Zhang J L, Zhang S J, Kong P P, Zhu J, Li X D, Liu J, Cao L Z, Jin C Q 2013 Superconductivity in copper intercalated topological compound $Cu_xBi_2Te_3$ induced via high pressure *Phys. C Supercond. its Appl.* **493**, 75–76

[6] Koski K J, Cha J J, Reed B W, Wessells C D, Kong D, and Cu Y 2012 High-Density Chemical Intercalation of Zero-Valent Copper into $Bi_2Se_3$ Nanoribbons. *J. Am. Chem. Soc.* **134**, 7584–7587

[7] Wang M and Koski K J 2016 Polytypic phase transitions in metal intercalated $Bi_2Se_3$. *J. Phys. Condens. Matter* **28**

[8] Wang Y L *et al.* 2011 Structural defects and electronic properties of the Cu-doped topological insulator $Bi_2Se_3$. *Phys. Rev. B* **84**, 075335

[9] López P A, Leal F M, and Derat R E 2016 Structural and Electronic Characterization of $Cu_xBi_2Se_3$. *J. Mex. Chem. Soc.* **60**, 101–107

[10] Koski K J, Wessells C D, Reed B W, Cha J J, Kong D and Cui Y 2012 Chemical intercalation of zerovalent metals into 2D layered $Bi_2Se_3$ nanoribbons. *J. Am. Chem. Soc.* **134**, 13773–13779

[11] Wilson J A, Disalvo F J and Mahajan S 1975 Charge-density waves and superlattices in metallic layered transition-metal dichalcogenides. *Adv. Phys* **24**, 117–201

[12] Rossnagel K 2011 On the origin of charge-density waves in select layered transition-metal dichalcogenides. *J. Phys. Condens. Matter* **23**, 101-107

[13] Matano K, Kriener M, Segawa K, Ando Y and Guo-qing Zheng 2016 Spin-rotation symmetry breaking in the superconducting state of $Cu_xBi_2Se_3$. *Nature Phys.* **12**, 852–854

[14] Ren Z, Taskin A A, Sasaki S, Segawa K and Ando Y 2010 Large bulk resistivity and surface quantum oscillations in the topological insulator $Bi_2Te_2Se$. *Phys. Rev. B* **82**, 241306

[15] Jia S, Ji H, Climent-Pascual E, Fuccillo M E, Charles M E, Xiong J, Ong N P and Cava R J 2011 Low-carrier-concentration crystals of the topological insulator $Bi_2Te_2Se$. *Phys. Rev. B* **84**, 235206

[16] Balakrishnan G, Baumberger F, Catlow C R A, Scanlon D O and Catlow P C R A 2012 Controlling bulk conductivity in topological insulators: Key role of anti-site defects. *Adv. Mater.* **24**, 2154-2158

[17] Richter W, Kokler H and Becker C R 1977 A Raman and Far-Infrared Investigation of Phonons in the Rhombohedral compounds. *Phys. Status Solidi* B **84**, 619–628

[18] Nakajima S J 1971 *Phys. D: Appl. Phys.* **4**, 685

[19] DiSalvo F J, Wilson J A, Bagley B G, and Waszczak J V 1975 Effects of doing on charge-density waves in layer compounds

[20] Lee M H, Chen C H, Tseng C M, Lue C S, Kuo Y K, Yang H D, and Chu M -W 2014 Concomitant charge-density-wave and Unit-cell-doubling structural transitions in $Dy_5Ir_4Si_{10}$. *Phys. Rev. B* **89**, 195142

[21] Hovden R et al. 2016 Atomic Lattice disorder in charge-density-wave phases of exfoliated dichalcogenides ($1T-TaS_2$)

[22] Wilson J A, Di Salvo F J, and Mahajan S 1974 Charge-Density Waves in Metallic, Layered, Transition-Metal Dichalcogenides. *Phys. Revs. Lett*. **32**, 882

[23] Williams P M, Scruby C, Clark W, Parry G 1976 CHARGE DENSITY WAVES IN THE LAYERED TRANSITION METAL DICHALCOGENIDES. J. Phys. Colloques. **37**, C4-139-C4-150.

[24] McMillan W L 1975 Landau theory of charge-density waves in transtion-metal dichalcogenides. *Phys. Rev. B* **12**, 1187

[25] DiSalvo F J, and Maurice Rice T 1979 Charge-density waves in transition-metal compounds. *Physics Today* **32**, 4, 32

[26] Huang F -T et al. 2012 Nonstoichiometric doping and Bi antisite defect in single crystal Bi2Se3 Phys. Rev. *B* **86**, 081104

[27] Kriener M, Seqawa K, Ren Z, Sasaki S, Wada S, Kuwabata S and Ando Y 2011 Electrochemical synthesis and superconducting phase diagram of $Cu_xBi_2Se_3$. *Phys. Rev. B* **84**, 054513

[28] Vaško A, Tichý L and Horák J 1974 Amphoteric nature of copper impurities in $Bi_2Se_3$ crystals. *Appl. Phys.* **5**, 217–221

[29] Tian Y, Osterhoudt G B, Jia S, Cava R J and Burch K S 2016 Local phonon mode in thermoelectric $Bi_2Te_2Se$ from charge neutral antisites. *Appl. Phys. Lett.* **108**

[30] Teyssier J, Homes C C, Akrap A and Lerch P 2012 Optical properties of $Bi_2Te_3Se$ at ambient and high pressure. *Phys. Rev. B* **86**, 235207

[31] Sokolov O B, Skipidarov S Y, Duvankov N I and Shabunina G G 2007 Phase relations and thermoelectric properties of alloys in the $Bi_2Te_3$-$Bi_2Se_3$ system. *Inorg. Mater.* **43**, 8–11

[32] Yuan J, Zhao M, Yu W, Lu Y, Chen C, Xu M, Li S, Loh K and Bao Q 2015 Raman Spectroscopy of Two-Dimensional $Bi_2Te_xSe_{3-x}$ Platelets Produced by Solvothermal Method. *Materials* **8**, 5007–5017

[33] Chen H *et al.* 2012 Phonon dynamics in $Cu_xBi_2Se_3$ (x=0, 0.1, 0.125) and $Bi_2Se_2$ crystals studied using femtosecond spectroscopy. *Appl. Phys. Lett.* **101**, *121912*

[34] Kumar A, Mishra V, Warshi M K, Sati A, Sagdeo A, Kumar R, Sagdeo P R 2019 Strain Induced Disordered Phonon Modes in Cr doped $PrFeO_3$. *J. Phys. Condens. Matter* **31**, 275602

[35] Chen C, Choe J & Morosan E 2016 Charge density waves in strongly correlated electron systems. *Rep. Prog. Phys.* **79**, 084505

[36] Zhu X, Cao Y, Zhang J, Plummer E W and Guo J 2015

Classification of charge density waves based on their nature. *Proc. Natl. Acad. Sci.* **112**, 2367–2371

[37] Johannes M D and Mazin I I 2008 Fermi surface nesting and the origin of charge density waves in metals. *Phys. Rev. B - Condens. Matter Mater. Phys.* **77**, 1–8

[38] Pouget J P 2016 The Peierls instability and charge density wave in one-dimensional electronic conductors. *Comptes Rendus Phys.* **17**, 332–356

[39] Wei M J, Lu W J, Xiao R C, Ly H Y, Tong P, Song W H and Sun Y P 2017 Manipulating charge density wave order in monolayer 1T -$TiSe_2$ by strain and charge doping : A first-principles investigation. *Phys. Rev. B* **96,** 165404

[40] Gannon L A Ph.D thesis. 2015 *Univ. Oxford*

[41] Kou C N, Shen D, Li B S, Quyen N N, Tzeng W Y, Luo C W, Wang M, Kuo Y K and Lue C S 2019 Characterization of the charge density wave transition and observation of the amplitude mode in $LaAuSb_2$. *Phys. Rev.* B **99**, 235121

[42] Chan S K and Heine V 1973 Spin density wave and soft phonon mode from nesting Fermi surfaces *J. Phys. F Met.Phys.* **3**, 795-809

[43] Rossnagel R K, Rotenberg E, Koh H, Smith N V and Kipp L 2005 Fermi surface, charge-density-wave gap, and kinks in $2H$-$TaSe_2$. *Phys. Rev. B - Condens. Matter Mater. Phys.* **72**, 1–4

[44] Whangbo M H and Canadell E 1992 Analogies between the Concepts of Molecular Chemistry and Solid-State Physics concerning Structural Instabilities. Electronic Origin of the Structural Modulations in Layered Transition-Metal Dichalcogenides. *J. Am. Chem. Soc.* **114**, 9587–9600